\documentclass{osa-article}
\usepackage{braket}
 \usepackage{color}
 \definecolor{Re}{rgb}{0.00,0.00,0.00}
\journal{oe}

\articletype{Research Article}

\begin{document}

\title{Classical simulation of high-dimensional entanglement by non-separable angular--radial modes}

\author{Shilong Liu,\authormark{1,2} Shikai Liu,\authormark{1,2} Chen Yang,\authormark{1,2} Zhaohuai Xu,\authormark{1,2} Yinhai Li,\authormark{1,2,3} Yan Li,\authormark{1,2} Zhiyuan Zhou,\authormark{1,2,3,*} Guangcan Guo,\authormark{1,2,3}  and Baosen Shi\authormark{1,2,3,4}}

\address{\authormark{1}Key Laboratory of Quantum Information, University of Science and Technology of China, Hefei, Anhui 230026, China\\
\authormark{2}Synergetic Innovation Center of Quantum Information \& Quantum Physics, University of Science and Technology of China, Hefei, Anhui 230026, China\\
\authormark{3}Heilongjiang Provincial Key Laboratory of Quantum Regulation and Control, Wang Da-Heng Collaborative Innovation Center, Harbin University of Science and Technology, Harbin 150080, China.}
\leftline{\authormark{4}drshi@ustc.edu.cn}

\email{\authormark{*}zyzhouphy@ustc.edu.cn}
\leftline{\homepage{http://www.quantumdrshi.com/}} 


\begin{abstract}
An analogous model system for high-dimensional quantum entanglement is
proposed, based on the angular and radial degrees of freedom of the improved Laguerre Gaussian mode. Experimentally, we observed strong violations of the Bell-CGLMP inequality for maximally non-separable states of dimension 2 through 10. The results for violations in classical non-separable state are in very good agreement with quantum instance, which illustrates that our scheme can be a useful platform to simulate high-dimensional non-local entanglement. Additionally, we found that the Bell measurements provide sufficient criteria for identifying mode separability in a high-dimensional space. Similar to the two-dimensional spin-orbit non-separable state, the proposed high-dimensional angular-radial non-separable state may provide promising applications for classical and quantum information processing.
\end{abstract}

\section{Introduction}
Non-local entanglement, a remarkable non-classical correction for spatially separated systems, is not only discussed in the context of fundamental physics, i.e., Einstein, Podolsky and Rosen (EPR) paradox \cite{Einstein1935} and Bell's inequality \cite{Clauser1978,Weihs1998,Collins2002}, but also widely applied to the fields of metrology \cite{Giovannetti2011}, quantum computation \cite{Kok2007} and communications \cite{Kimble2008,Gisin2002,Bouwmeester1997}.  Moreover, there exists a type of entanglement-like state, sometimes named as local entanglement between different degrees of freedom (DoFs) in a single system\cite{Spreeuw1998,Toeppel2014}. In the literature, this phenomenon is variously called  'intra-system entanglement', 'nonquantum entanglement', 'hybrid entanglement' or 'non-separable state'\cite{Toeppel2014,toninelli2019concepts,doi:10.1080/00107514.2019.1580433}.
  It may also appear in classical systems, i.e., a classical nonseparable optical mode \cite{aiello2015quantum,Souza2007,lee2002experimental}; and in single particle \cite{Hasegawa2003,Zhou2016,Fickler2012,Karimi2010}.
  'Nonlocality' is the feature distinguishing non-local entanglement from local entanglements.  Recently, some significant advances suggest that the 'Nonlocality' is not a necessary condition for implementing many quantum computing tasks \cite{Borges2010}, for example, quantum walk \cite{Goyal2013,Cardano2015} and several parallel-search algorithms \cite{Perez-Garcia2018,DeOliveira2005}, where local non-separable state can increase computation resources while keeping the physical number of particles constant.  Also, because of the equivalent mathematical form between them,  the non-separable state supports an effective platform to simulate the behaviors of quantum non-local entanglements \cite{aiello2015quantum,lee2002experimental,Spreeuw2001}, i.e., researches of the Bell measurement \cite{Borges2010,Kagalwala2013} and quantum contextuality \cite{Passos2018,Karimi2010}.  Classical analogy can help one to understand and visualize the behaviors of a 'true' quantum state in quantum world \cite{dragoman2013quantum}.

One well-known non-separable concept is the spin-orbit state \cite{Souza2007,Ndagano2017,Borges2010,
 Karimi2015,Kagalwala2013,doi:10.1080/00107514.2019.1580433}. It has been widely studied in applications in optical metrology, sensing \cite{Toeppel2014,berg2015classically}, communication protocols \cite{Ndagano2017,barreiro2008beating}, and key devices for quantum computation, for example, CNOT gates \cite{DeOliveira2005,marrucci2011spin}. Experimentally, a convenient way to realize non-separable modes is to employ spin angular momentum (SAM) and orbital angular momentum (OAM) DoFs of light, i.e., $\left| H \right\rangle \left| L \right\rangle  + \left| V \right\rangle \left| { - L} \right\rangle $. However, the SAM has only two orthogonal eigenstates, and therefore the spin-orbital non-separable state has poor dimensional scalability.
 In contrast, the Laguerre-Gaussian modes ($LG_L^P \to \left| L \right\rangle \left| P \right\rangle $) span an infinite dimensional space in both angular $L$ and radial mode $P$ indices, and therefore the potential to construct high-dimensional (HD) non-separable state with angular--radial number \cite{Salakhutdinov2012,loffler2012hybrid}.

 Over the past few decades, the angular mode has drawn much attention \cite{Allen1992,Dada2011,Mair2001,Yao2011} whereas  the radial mode is not very attractive because of  difficulties in sorting and detection. There has been some progress in radial mode through in both classical and quantum fields \cite{Karimi2014HOM,Krenn2017,Salakhutdinov2012,Trichili2016,Zhang2014,Karimi2014,Zhang2018}, specifically, exploring full-field quantum corrections \cite{Salakhutdinov2012,Krenn2014}. Very recently, several significant advances have been made in sorting radial modes using the accumulated Gouy phases \cite{Gu2018,Zhou2017}, and measuring full fields in a technique that exploits intensity- flattening \cite{Bouchard2018}
 . These latest technological advances indicate that it is possible to build a HD non-separable angular--radial state $\left| {{L_1}} \right\rangle \left| {{P_1}} \right\rangle  + \left| {{L_2}} \right\rangle \left| {{P_2}} \right\rangle  + ...$ {\color{Re}{to go beyond the existing spin-orbital non-separable state in dimension.}}

    In this article, we construct a HD angular--radial non-separable state (HD-ARNS) to simulate the HD quantum entanglement using the classical revised LG modes, where the HD-ARNS has the form of a maximally entangled state (MHD-ARNS):  $\sum\nolimits_{j = 0}^{d - 1} {{{\left| j \right\rangle }_L}{{\left| j \right\rangle }_P}} $. Experimentally, we observe the violation of the Bell-CGLMP-like inequality  from d=2 to 10. The results for the violation of MHD-ARNS are in very good agreement with quantum situations described in \cite{Dada2011}, and therefore illustrates that our scheme forms a useful platform to simulate classically various HD entanglement of two particles \cite{Dada2011,Mair2001,Liu20183DMES,Kues2017}.

 For non-local entanglement, violations of the Bell inequality demonstrate that the corresponding quantum state cannot be explained by local hidden-variable theory (LHV). For local non-separable states, the violation of Bell-like inequality is an effective quantitative tool in studying classical optical coherence and mode (non-) separability \cite{Souza2007,Borges2010,Kagalwala2013}. However, there has been very little experimental research because of  difficult manipulation. {\color{Re}{Very recently, attempts have been made to simulate the HD entanglement by measuring the difference of light intensity at two output ports of a Mach-Zehnder (MZ) interferometer \cite{Li2018}}}. However, increasing the dimensions has been difficult because of the complicated experimental setup and the low value of the Bell-like inequality \cite{Li2018}. In our works, to verify mode non-separability of the generated angular--radial modes, we demonstrated a dynamic process from  separation to nonseparation for a three-dimensional angular--radial non-separable mode, for which the distributions of the Bell-CGLMP-like inequality and interference visibility were analyzed in detail.

 \begin{figure}[htbp]
  \centering
  \includegraphics[width=10cm]{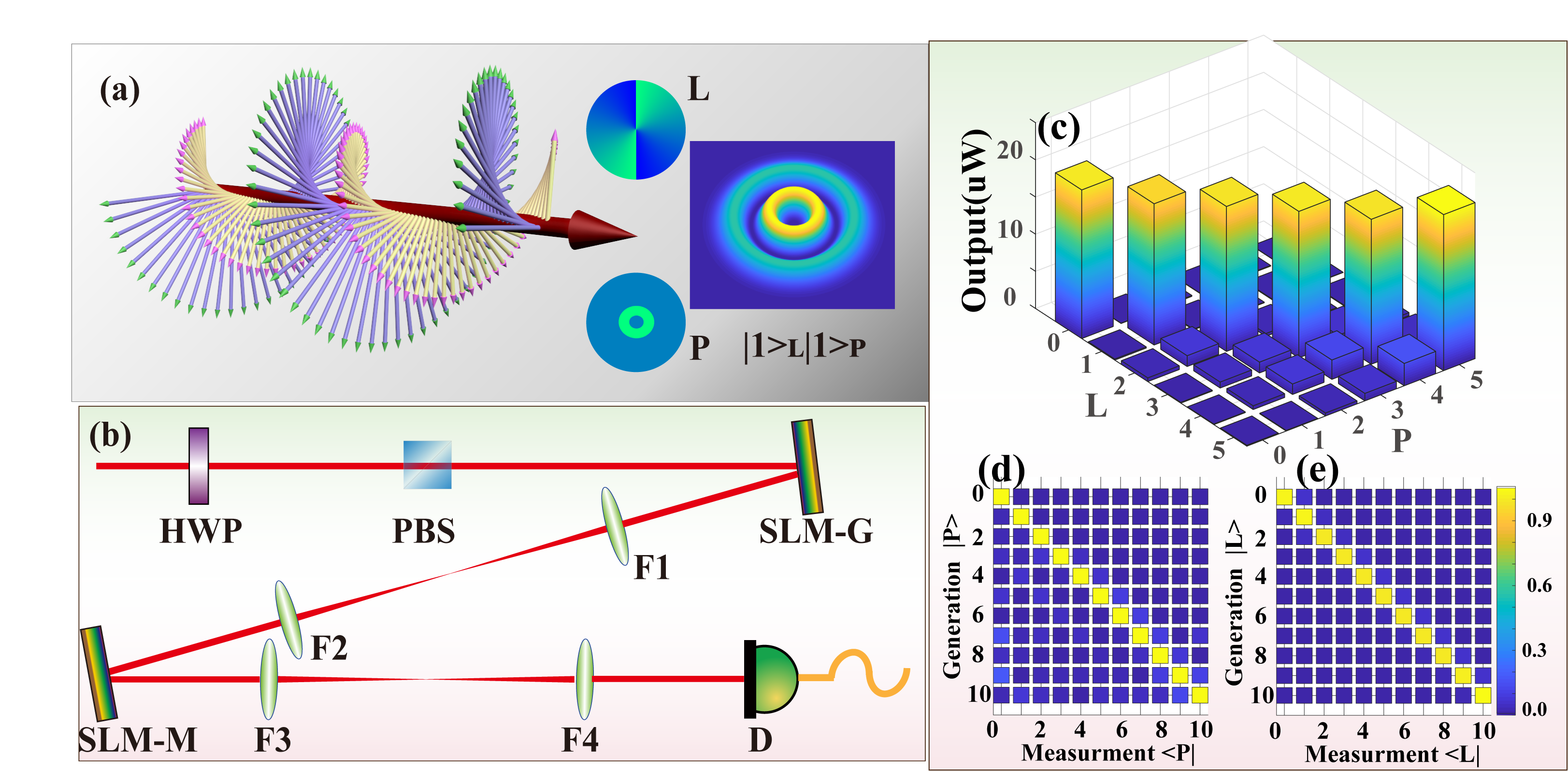}
  \caption{Setups for the generation of a HD-ARNS and measured modal decomposition density: (a) spatial vector distribution of the spatial vector, phase, and intensity for LG beam $\ket{1}_L\ket{1}_P$; (b) optical layout for generation and detection HD-ARNS. The input laser source is a semiconductor laser at the 780 nm. HWP: half wave plate. PBS: polarization beam splitter. F1-4: convex lens of two inches with $f$=300 mm. D: a power meter for detection. (c): a six-dimensional ARNS. (d) and (e) show the situation of a pure LG mode when radial L=0 and angular P=0, respectively; the range in each testing is from 0 to 10; the normalized distribution is calculated by the corresponding each row. The color scale represents the powers obtained by normalization of each row.  }
\end{figure}

\section{Principle}
{\color{Re}{ The electric field of the LG beam $LG_{L{\rm{ = 1}}}^{P{\rm{ = }}1}$ with both angular $L$ and radial $P$ modes is given as $E_{L,P}(r,\phi,z)$. Because we want to simulate the biphoton MHD-MES: $\sum\nolimits_{j = 0}^{d - 1} {{{\left| j \right\rangle }_1}{{\left| j \right\rangle }_2}} $, we'd like to use the Dirac notation to describe a classical LG mode for simplicity: $E_{L=j,P=j}(r,\phi,z)$->${\left| j \right\rangle _L}{\left| j \right\rangle _P}$.}}  When both  are zero, the beam degenerates to a conventional Gaussian beam. Fig.~1(a) shows the spatial vector and intensity distribution of the state ${\left| 1 \right\rangle _L}{\left| 1 \right\rangle _P}$ from a three-dimensional perspective. The angular number $L$ gives rise to a helical wave front, which affords a representation of light as carrying a well-defined orbital angular momentum; the radial number $P$ indicates there are $P+1$ concentric rings on the wavefront and intensity pattern, suggesting a charge with hyperbolic momentum charge \cite{Plick2015}. The manipulation of the $P$ mode is experimentally more difficult than the $L$ mode, where the overlap between different radial modes is strongly dependent on the beam waist of the basis, the propagation distance, and collection systems  \cite{Karimi2014HOM,Salakhutdinov2012,Plick2015}. The development of amplitude-phase encoding technology \cite{Bolduc2013} and demagnification- regime \cite{Bouchard2018,Salakhutdinov2012,Karimi2014HOM} overcomes these difficulties, although at present the reflection efficiency is low \cite{Karimi2014HOM, Sephton2016}. Different from the previous regimes, we employ a revised LG mode for manipulating non-separable state with both DoFs, where the beams for all the basic states have equal size, but different waists ${w_0}(L,P) = {w_0}\left( {0,0} \right)/\sqrt {|L| + 2P + 1} $, and orthogonality is retained for both DoFs when we consider the Gaussian mode inner production of the single mode fiber \cite{vallone2017role,Roux2014} (also see in appendix). The regimes of revised LG mode have many unique advances. On the one hand, the equally 4$f$ system before single mode fiber [F3 and F4 in Fig. 1(b)] ensures the high efficiency in collection, which is beneficial to be expanded in single photon level. On the other hand, one can avoid  higher radial modes exceeding the periphery of the SLMs (details in appendix), which ensures the generation can carry more transverse modes. Recently, the revised LG mode has shown many outstanding abilities in multi-mode superposition and quantum-key-distribution for diminishing atmospheric turbulence \cite{ndagano2018characterization,Liu2018Cat}.

    Using an amplitude-phase encoding hologram, we successfully generate and detect the full field with both DoFs [Fig.~1(b)]. A spatial light modulator, labeled SLM-G, is used to generate the angular--radial non-separable state $\ket{L}\ket{P}$. The beam is exactly imaged onto the surface of  another SLM-M for measurement by a 4$f$ system. The beam is coupled to a single mode fiber by another 4$f$ lens. The coupling efficiency for Gaussian mode is equal to 75\%.
 \section{Results}
 \subsection{Spatial modal decomposition}
The results obtained by manipulating the HD-ARNS are shown in Fig. 1. Fig. 1(c) depicts the modal decomposition density for a six-dimensional ARNS ${\left| \psi  \right\rangle _{d = 6}} = \frac{1}{{\sqrt 6 }}\sum\limits_{j = 0}^5 {{{\left| j \right\rangle }_L}{{\left| j \right\rangle }_P}} $, where one needs to make 36 ($d^2$) projection measurements $\left\{ {\left\langle \varphi  \right|} \right\}{\rm{ = }}\left\{ {{{\left\langle m \right|}_L}{{\left\langle j \right|}_P}} \right\}$. {\color{Re}{Each of projected basis, i.e., $ {{{\left\langle m \right|}_L}{{\left\langle j \right|}_P}}$, represents a conjugate state of angular mode $L$ and radial mode $P$. Experimentally, we load the phase of the projected basis onto the  SLM-M (see Fig. 1(b)) to realize the projection measurements: ${\left\langle m \right|_L}{\left\langle j \right|_P} \cdot {\left| n \right\rangle _L}{\left| q \right\rangle _P} = {\delta _{m,n}} \cdot {\delta _{j,q}}$ \cite{Gibson2004,Yao2011}, where we employ the  amplitude-encoded technology to generate and measure the corresponding states \cite{Bolduc2013,Liu2018Cat}}}.

For characterizing the crosstalk, one defines the power-visibility $V = \sum\nolimits_i {{I_{ii}}} /\sum\nolimits_{ij} {{I_{ij}}} $, where $I_{ij}$ represents the obtained average power ${\left| {\left\langle {\varphi {\rm{|}}\psi } \right\rangle } \right|^2}$. The power-visibility is 87.26\%$\pm$0.03 for a six-dimensional ARNS, where the error  comes from the jitter in power; it is a standard deviation, estimated from simulations that each data is assumed to follow the Poisson's distribution. The power-visibility will decrease for higher-dimensional state, i.e., 77.6\% for a ten-dimensional ARNS. We measured the normalized modal decomposition density for single states with one DoF [Fig. 1(d) and 1(e)]. The power visibilities calculated from two matrices are 92.78\% and 82.9\% for the angular and radial modes, respectively, where the visibility of the radial mode is quite low owing to imperfect overlaps. Nevertheless, the large diagonal elements and small cross-talk indicate that the states generated are strongly orthogonal and of high quality.

The generation of MHD-ARNS in $d$ dimensions is written ${\left| \psi  \right\rangle _d} = \frac{1}{{\sqrt d }}\sum\limits_{j = 0}^{d - 1} {{{\left| j \right\rangle }_L}{{\left| j \right\rangle }_P}} $ , which has the same mathematical form as the two particles \cite{Collins2002,Dada2011}. We find that the HD Bell-CGLMP inequality  also holds for the MHD-ARNS, i.e., $S_d>2$,  where ${S_d} = \sum\limits_{k = 0}^{[d/2] - 1} {{S_k}} ({A_a},{B_b})$ is the Bell-CGLMP expression. In our system, the observations of A $\theta _L^a = \frac{{2\pi }}{d}\left[ v + a/2\right]$  and B $\theta _P^b = \frac{{2\pi }}{d}\left[ { -w + 1/4{{( - 1)}^b}} \right]$ are angular and radial DoFs, respectively, and two labels of $a$ and $b$ have discrete values 0 and 1. Therefore, the measurement bases for the two corresponding DoFs may be defined as:
\begin{equation}
\begin{split}
     \left| \theta  \right\rangle _L^a \otimes \left| \theta  \right\rangle _P^b =  \frac{1}{{\sqrt d }}\sum\limits_{j = 0}^{j = d - 1} {\exp (i\theta _L^aj){{\left| j \right\rangle }_L}}  \otimes \frac{1}{{\sqrt d }}\sum\limits_{j = 0}^{j = d - 1} {\exp (i\theta _P^bj){{\left| j \right\rangle }_P}}
\end{split}
\end{equation}
The expression for the intensity for two DoFs in a joint measurement is equal to the joint probabilities, obtained in the original paper \cite{Collins2002}
\begin{equation}
\begin{split}
     I\left( {\theta _L^a,\theta _P^b} \right){\rm{ = }} \frac{1}{{{d^{3/2}}}}{\left| {\sum\limits_{j = 0}^{j = d - 1} {\exp \left( {i\frac{{2\pi }}{d}j\left( {v - w + a/2 + {{\left( { - 1} \right)}^b}/4} \right)} \right)} } \right|^2}
\end{split}
\end{equation}

 \subsection{Bell-CGLMP-like  inequality for high-dimensional non-separable state }
 The key results for this paper are the HD Bell-CGLMP-like interference curves and inequalities [Fig. 2]. For a two dimensional state ${\left| 0 \right\rangle _L}{\left| 0 \right\rangle _P} + {\left| 1 \right\rangle _L}{\left| 1 \right\rangle _P}$, the measured intensity is similar to the coincidences of two-dimensional biphoton entanglement state \cite{Jack2010,kim2006phase}. Two experimental interference datasets plotted in Fig. 2(a) were obtained by changing the value of the radial phase ${\theta _P}$ while fixing the value of the angular phase ${\theta _L}$ to 0 and $\pi$. The interference data for the ten-dimensional ARNS [ Fig. 2(b) ] was obtained by setting the radial phase to a constant value.{\color{Re} The theoretical fits to data  are plotted as solid lines. The error bars represent $ \pm 1$ standard deviation, which were estimated from statistical simulations that the data were assumed to follow a Poisson's distribution \cite{hannam1999estimating}. In our system, we found that the jitter distribution of input laser power is similar to a Poisson distribution. Therefore, the assumption of  all the data follows the Poisson distribution is reasonable in simulations.}

 As the optical system is imperfect, some basic phase appears between low-order and high-order modes \cite{Liu20183DMES}; the interference curve appears to have shifted to the left [Fig. 2(b)], especially for the higher-dimensional states. The shift gives rise to a dislocation between sample and background (see Fig. 4(a) in \cite{Liu20183DMES}). To overcome the dislocation, one needs to increase the basic phase slightly for both radial and angular parts in Eq. (1).

 Fig.~2(c) shows the values of the Bell-CGLMP expression  $S_d$ versus the dimension $d$ measured by two-photon HD entanglement (blue bars) with the entanglement concentration given in \cite{Dada2011}, MHD-ARNS (red bars) in our setups. The blue graduated area ${S_d} \le 2$ shows that this state satisfies the LHV theory, and the green line marks the upper bound for violations of the maximally HD entanglement states. {\color{Re}{ In our systems, the dimension in which the Bell inequality is violated for MHD-ARNS goes up to 10; the value obtained, ${S_{d{\rm{ = 10}}}}{\rm{ = }}2.650 \pm 0.035$ is violated 18.6 standard deviations as the classical bound, i.e., $S_{d=10}=2$.} } An interesting feature is that the maximum limit for maximally HD entangled states is greater than $2\sqrt 2 $ for two dimensions. In our system, we find  violations of ${S_{d{\rm{ = 4}}}}{\rm{ = }}2.883 \pm 0.017$ and ${S_{d{\rm{ = 5}}}}{\rm{ = }}2.912 \pm 0.018$ in four and five dimensions, where the standard deviations violation are 3.2 and 4.6 to beyond $\sqrt{2}$, respectively. Because of the crosstalk between both $L$ and $P$ modes, the violations are weaker as dimension increases. Both local and non-local entangled states can mathematically go beyond a kind of limit, i.e., ${S_d}{\rm{ = }}2$, although there are different fundamental physical meanings. The results for violations of MHD-ARNS are in very good agreement with quantum instance  \cite{Dada2011}, which illustrates that our scheme may be a useful platform to simulate classically HD non-local entanglement.
\begin{figure}[htbp]
    \centering
    \includegraphics[width=12cm]{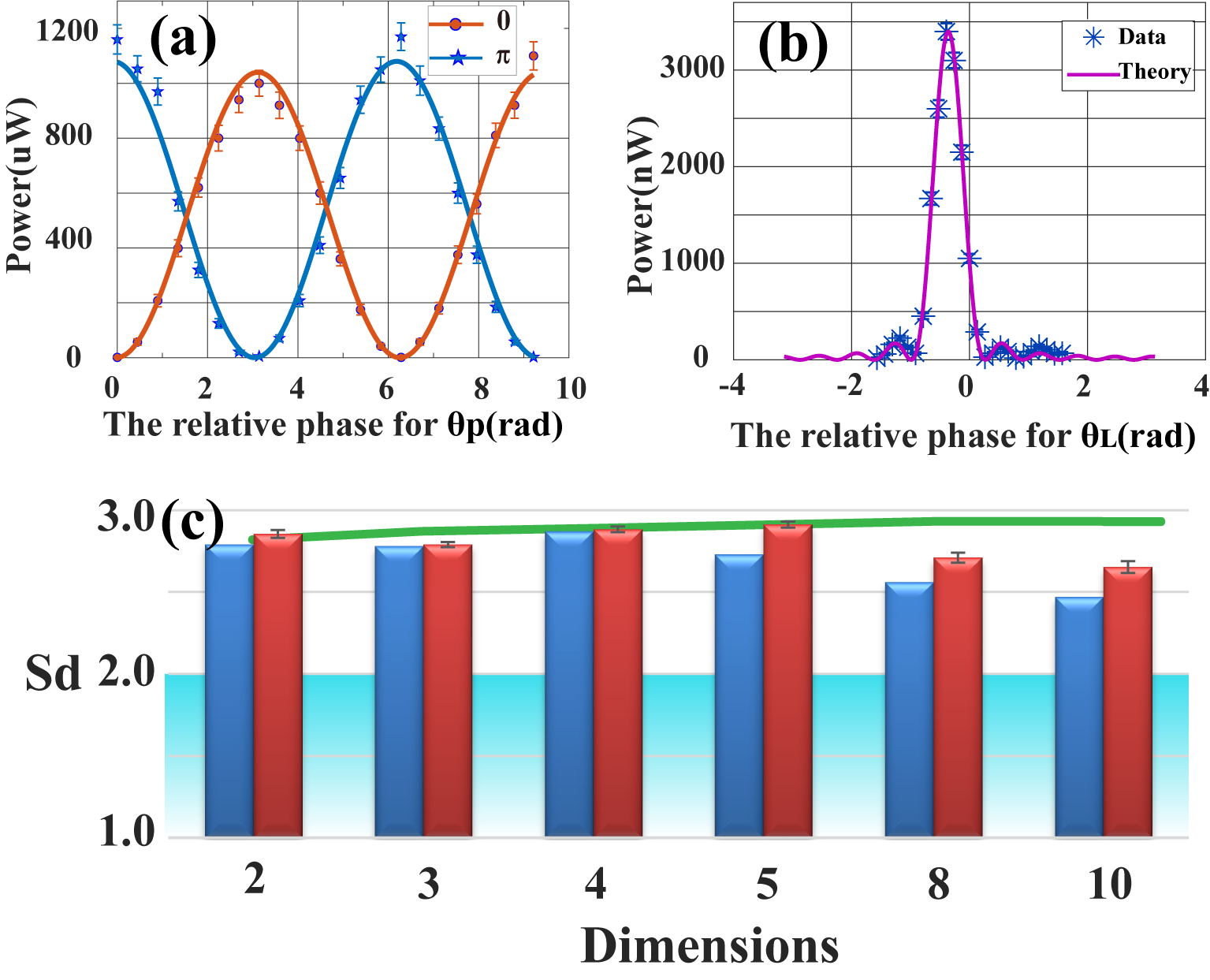}
    \caption{Results for Bell-CGLMP-type interference curves and the value of Bell-CGLMP-like inequality. (a) and (b) show the interference curves for the MHD-ARNS in d=2 and d=10, respectively. (c) shows the violations of the Bell-CGLMP-like inequality for nonlocal HD entangled states (blue) obtained in \cite{Dada2011} and for local HD non-separable states (red) measured in our system. }
  \end{figure}
  \begin{figure}[hbt]
  \centering
  \includegraphics[width=12cm]{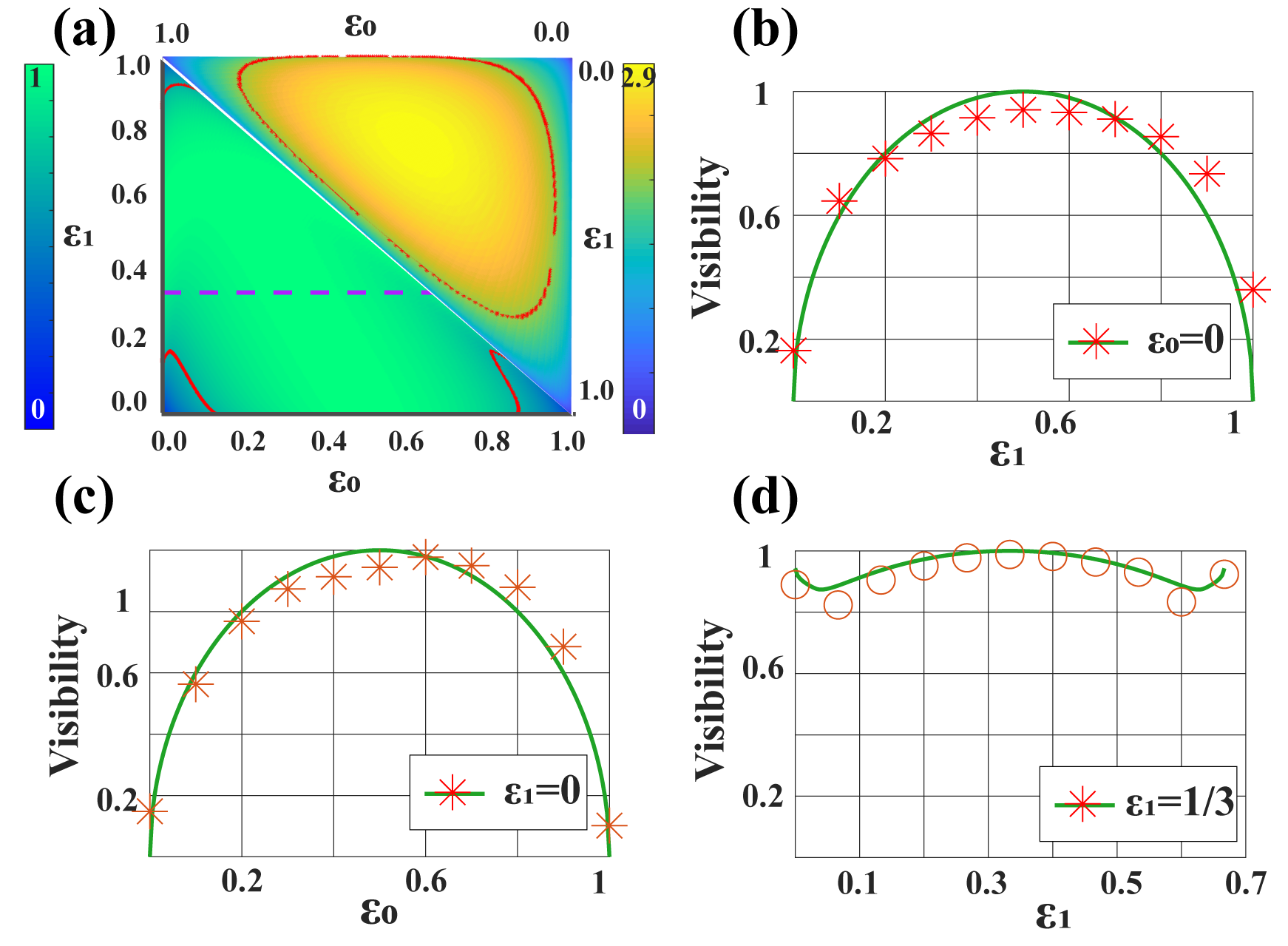}
  \caption{
    Theoretical and experimental distributions of values of the Bell-CGLMP expression $S$ and interference visibility $V$ for a three-dimensional ARNS. (a) A three-dimensional surface shows the $S$ and $V$ plotted using two parameters ${\varepsilon _0}$ and ${\varepsilon_1 }$; the purple dashed line marks the situation for ${\varepsilon_1}{\rm{ = }}1/3$; the red boundary marks instance with $S_{d=3}=2$ and 71\%. (b) Vertical distributions of visibility versus ${\varepsilon_0}$ with ${\varepsilon _0}{\rm{ = }}1/3$; (c) and (d) are instances of visibility changing with ${\varepsilon _0}$n, where ${\varepsilon_1}=0$ and ${\varepsilon_1}=1/3$, respectively.
  }
\end{figure}
  \subsection{The criterion for mode (non)separability}
For the HD non-local quantum entanglement, violation of the Bell-CGLMP inequality provides an effective criterion indicating the existence of non-locality, as well as a valid criterion for  (non-)separability of the HD local entanglement. {\color{Re} Also, one can evaluate the separability by the visibility measured from Bell-CGLMP-type interference data. \cite{Souza2007}.  In two dimensional system,  the area for values of Bell-like expression beyond one special boundary, i.e., ${S_2} \ge 2$, is equal to the visibility, i.e.,   $V \ge 71\%$ \cite{Souza2007}.  The visibility shows unique advantage as a useful criterion for mode (non-)separability  because the number of measurements is less than in Bell-CGLMP inequality. Experimentally, many relative works have been demonstrated in two-dimensional classical entanglement systems \cite{Souza2007,Borges2010,Kagalwala2013}.

It is natural to ask a question: are the boundary of visibility and Bell-CGLMP inequality the same for HD non-separable state?
Here, we expand the results to classical HD non-separable mode. Consider a three-dimensional ARNS,}
\begin{equation}
  {\left| \varphi  \right\rangle _3} = \sqrt {{\varepsilon _0}} {\left| 0 \right\rangle _L}{\left| 0 \right\rangle _P} + \sqrt {{\varepsilon _1}} {\left| 1 \right\rangle _L}{\left| 1 \right\rangle _P} + \sqrt {1 - {\varepsilon _0} - {\varepsilon _1}} {\left| 2 \right\rangle _L}{\left| 2 \right\rangle _P}
\end{equation}
 where ${\varepsilon _0}$ and ${\varepsilon_1}$ are two real coefficients for the states $\ket{0}_L\ket{0}_P$ and  $\ket{1}_L\ket{1}_P$, a constraint ${\varepsilon _0} + {\varepsilon _1} \le 1$ should be acquired. The theoretical simulation of $S$ versus two coefficients are plotted [Fig. 3(a), right top], where the red dots mark the limit $\left| S \right| = 2$. Furthermore, one can evaluate the visibility $V\left( {{\varepsilon _0}\;,{\varepsilon _1}} \right)= {I_{Max}} - {I_{Min}}/{I_{Max}} + {I_{Min}}$ from the  joint correction intensity power $I\left( {{\varepsilon _0}\;,{\varepsilon _1}} \right){\rm{ = }}{\left| {{{\left( {\left| \theta  \right\rangle _L^a \otimes \left| \theta  \right\rangle _P^b} \right)}^ {\dag}} \cdot {{\left| \varphi  \right\rangle }_{d = 3}}} \right|^2}$. Similarly, theoretical simulations of visibility versus two parameters were plotted [Fig. 3(a), left bottom]. Experimentally, one can change their value in the acquired phase hologram. When  ${\varepsilon _0}$ or ${\varepsilon_1}$  is equal to zero, the dimension of the MHD-ARNS diminishes from three to two. Fig. 3(b) and 3(c) present the corresponding dynamic processes, where the visibility (state) runs from minimum (separable) to maximum (non-separable) and to minimum (separable) as the other parameter increases. This behavior is  similar to the previous demonstration in spin-orbital system \cite{Souza2007}. For the nonzero situation, i.e., ${\varepsilon _1}{\rm{ = }}1/3$, the state is non-separable regardless of the value of ${\varepsilon _0}$, where high visibility is maintained [Fig. 3(d)].

 Interestingly, the distributions of visibility are different from the value given by the Bell-CGLMP-like expression. For example, in a three-dimensional ARNS, the area for values of Bell-CGLMP-like expression beyond the one special boundary, i.e., ${S_d} \ge 2$ is smaller than the visibility-value, i.e., $V \ge 71\% $ [see red boundary in Fig. 3(a)], which indicates that the criterion for (non-)separability using the inequality values is more rigorous than for visibility. Due to the same mathematical forms between classical or quantum HD entanglement, we believe the demonstrated relation is also suitable to the non-local entanglements.

\section{Discussion}
  Using classical LG waves with both angular and radius DOFs, we construct a HD non-separable modes to simulate the quantum high-dimensional entanglements. Experimentally, we observe  the violation of a Bell-CGLMP-like inequality for HD non-separable modes with both angular and radial DoFs. On the one hand, the violations demonstrated in a classical system were compared with the non-local two-photon HD entangled state, and demonstrated that our regime provides an effective platform to simulate non-local HD entanglements classically. On the other hand, it is an effective criterion for mode (non-) separability using HD Bell's measurement. Some significant potential applications of the generations are to be developed in future.
\begin{itemize}
\item{i)} The concepts are easily expandable to single photon level by changing the inputs, for example, the construction of $d$-level four-particle cluster states by two photons; recently, these cluster states were prepared using two photons in both the time and frequency domains \cite{Reimer2018}.
\item{ii)} For the CGLMP inequality, the maximal violation is not the situation of the maximally entangled state \cite{acin2002quantum}; one can test it by changing the input state. Also,the HD maximally non-separable state supports an effective method to test or mimic noncontextuality hidden--variable models in higher dimensional space, i.e., the Kochen--Specker theorem in HD system \cite{Passos2018,Cabello1996}, and to study violations of other types of Bell-inequalities, such as the Son--Lee--Kim inequality \cite{son2006generic,datta2017measuring}.
\item{iii)} One can generate the group of HD angular--radial Bell states in single particle with high fidelity, where the non-diagonal state and phase between the entangled modes are easily manipulated compared with the non-local Bell state in two photons \cite{Liu20183DMES}. The corresponding Bell states could be used to high dimensional quantum computation \cite{wang2017generation,Babazadeh2017}.
\item{iv)} Recently, some effective proposals to quantify entanglement dimensionality for biphoton entanglement have been proposed, namely, the TILT basis \cite{bavaresco2018measurements,friis2018entanglement}. We believe it can be tested in our proposed maximally angular--radial non-separable mode.
\end{itemize}
  From the viewpoint of a computing resource, this type of DoF entanglement $\sum\limits_{j = 0}^{d - 1} {{{\left| j \right\rangle }_L}{{\left| j \right\rangle }_P}} $ in a single system (particle) is equal to the various already prepared HD quantum entanglement of two particles. Compared with the entanglement between multiple photon in non-local systems, the entanglement between DoFs in local systems is well controlled, has a higher detection rate, and is strongly robust to noise, which enables several special computation protocols to be implemented with high efficient.

\section*{Appendix}
\subsection*{ The fundamental characteristics of the high-dimensional non-separable angular--radial state}
    The normalized OAM eigenstate associated with the Laguerre Gaussian (LG) mode can be represented as $\left\langle {Cylin(r,\phi ,z)} \right.\left| {L\;,P} \right\rangle  \to LG_P^L(r ,\phi ,z)$ in the cylindrical representation. The electric field is given by \cite{Allen1992,Yao2011}.
   \begin{equation}
\begin{array}{l}
\left\langle {Cylin(r,\phi ,z)} \right.\left| {L\;,P} \right\rangle {\rm{ = }}\sqrt {\frac{{2P!}}{{\pi (P + \left| L \right|)!}}} \frac{1}{{w(z)}}{\left( {\frac{{\sqrt 2 r}}{{w(z)}}} \right)^{\left| L \right|}} \cdot L_P^{|L|}\left( {\frac{{2{r^2}}}{{{w^2}(z)}}} \right) \cdot \exp \left( { - \frac{{{r^2}}}{{w{{(z)}^2}}}} \right)\\
\;\;\;\;\;\;\;\;\;\;\;\;\;\;\;\;\;\;\;\;\;\;\;\;\;\;\; \;\;\;\;\;\times \exp \left( {i\left( {L\phi  - \frac{{k{r^2}z}}{{2({z^2} + z_r^2)}} + \left( {2P + \left| L \right| + 1} \right)a\tan \left( {\frac{z}{{{z_r}}}} \right)} \right)} \right)
\end{array}
\end{equation}

Where $1/e$ radius of the Gaussian term is given by $w(z) = w(0)\sqrt {1 + {{\left( {z/{z_r}} \right)}^2}} $ with $w(0)$ being the beam waist; $L_P^{|L|}(x)$ is the associated generalized Laguerre Gaussian polynomial;  $\left( {2P + \left| L \right| + 1} \right)a\tan \left( {z/{z_r}} \right)$ gives the Gouy phase.

   The LG modes have infinite dimensions in both angular $L$ and radial $P$ number, and meet the corresponding orthogonality:
   \begin{equation}
     \int\limits_0^\infty  {\int\limits_0^{2\pi } {LG_L^P{{(r,\phi )}^*}} }  \cdot LG_{L'}^{P'}(r,\phi )rdrd\phi  = {\delta _{ - L,L'}}{\delta _{P,P'}}
   \end{equation}
Where the orthogonality of angular mode $L$ is from the angular integral:
\begin{equation}
  \int\limits_0^{2\pi } {\exp (i(L - L')\phi )d\phi  = 2\pi *} {\delta _{L - L',0}}
\end{equation}

 And the orthogonality of radial mode $P$ comes from the orthogonality between generalized Laguerre Polynomial:
 \begin{equation}
   \int\limits_0^\infty  {L_P^{|a|}{{(r)}^*}L_{P'}^{|a|}(r)dr = \frac{{\Gamma (a + P + 1)}}{{P!}}} {\delta _{P,P'}}
 \end{equation}
  These two variables are independent in mode orthogonality, which illustrates that one can manipulate them independently.

  \begin{figure}[htp]
  \centering
  \includegraphics[width=10cm]{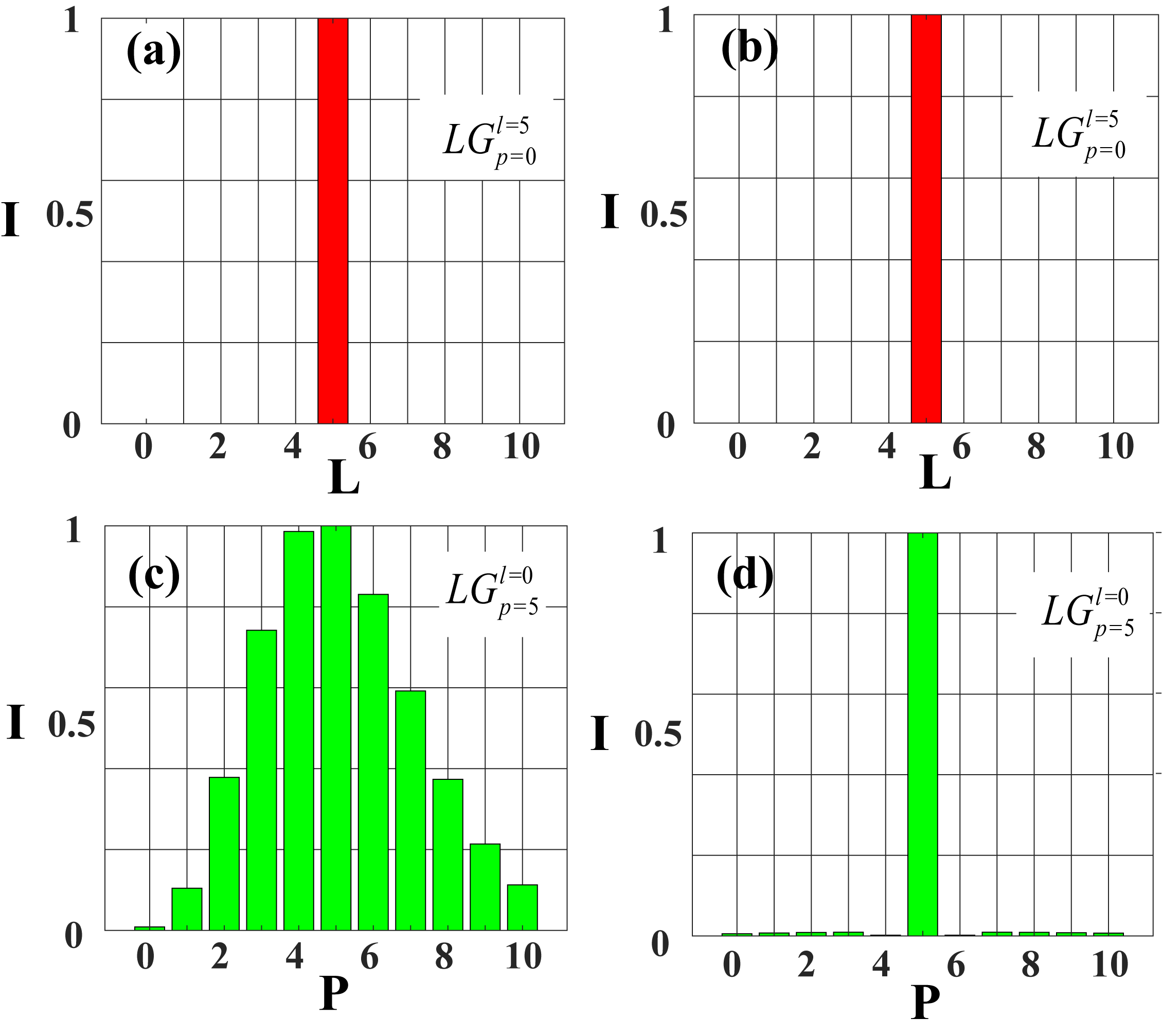}
  \caption{The orthogonality for angular and radial LG modes. (a) and (c): The orthogonality of $LG_0^5$ and $LG_5^0$  for standard LG modes. (c) and (d): The corresponding orthogonality for the revised LG modes. In the simulation, the beam waist $w_0$ for input Gaussian beam is 1000 um, and the mode size $W$ on the backward SLM is equal to 750 um. }
\end{figure}

  For detecting LG modes, we employ a 4-$f$ system. In this scheme, we need to consider the overlap integral between LG mode and single mode fiber (SMF). In that case, the orthogonality of radius mode will not meet because of the normalized Gaussian mode of the SMF (GM-SMF) \cite{Roux2014,Bouchard2018}:
  \begin{equation}
    G(u,w) = \frac{1}{w}\sqrt {\frac{2}{\pi }} \exp \left( { - \frac{{|u{|^2}}}{{{w^2}}}} \right)
  \end{equation}
The $u$ represents the transverse coordinates on the end of the SMF; $w$ is the size of mode on the end of the SMF. For reducing the effects from the Gaussian inner product, an alternative approach is used to expand the size of the Gaussian mode on the SLM by increasing the (de) magnification between the SLM and the SMF. In other words, one can reduce the distributions of modes on the end of SMF by designing a telescope system \cite{Bouchard2018}.

However, the drawback appears with the low coupling efficiency, which can be suppressed by the order of magnification. Besides, one can manipulate the beam waist while keeping the equally beam size for different high order modes, which is mathematically equal to the demagnification mentioned above by a group of the telescope lens. The latter method has a considerable advantage that the beam size of hologram with amplitude and phase is not beyond SLMs. As the LG mode becomes larger, the divergence angle $ \sim \sqrt {|L| + 2P + 1} $ increases drastically. For example, for the LG mode $LG_{P= 10}^{L= 10}$, the beam size will be expanded to 5.56 times than Gaussian beam. While in our regime, the beam sizes for all the modes are equal. What the method used here is to change the beam waist ${w_0}(L,P){\rm{ = }}{w_0}(0,0)/\sqrt {|L| + 2P + 1} $ for different LG mode. In that case, the orthogonality for angular mode still hold. However, it does not meet for the radial modes \cite{vallone2017role}. Nevertheless, the orthogonality for both angular and radial modes will hold by considering the Gaussian integral of SMF in Eq. (8), where the overlap is  given by the following equation \cite{Roux2014}:

\begin{equation}
\begin{array}{l}
\left\langle {LG_L^P|LG_{L'}^{P'}} \right\rangle  =
\int\limits_0^\infty  {\int\limits_0^{2\pi } {LG_L^P{{(r,\phi )}^*}} }  \cdot LG_{L'}^{P'}(r,\phi )*G(r,W)rdrd\phi
\end{array}
\end{equation}

Where $W$ is the size of the mode when the beam is imaged (magnified) backward onto the SLM, it is associated with the mode size $u$ in Eq. (8).
\begin{figure}[htp]
  \centering
  \includegraphics[width=10cm]{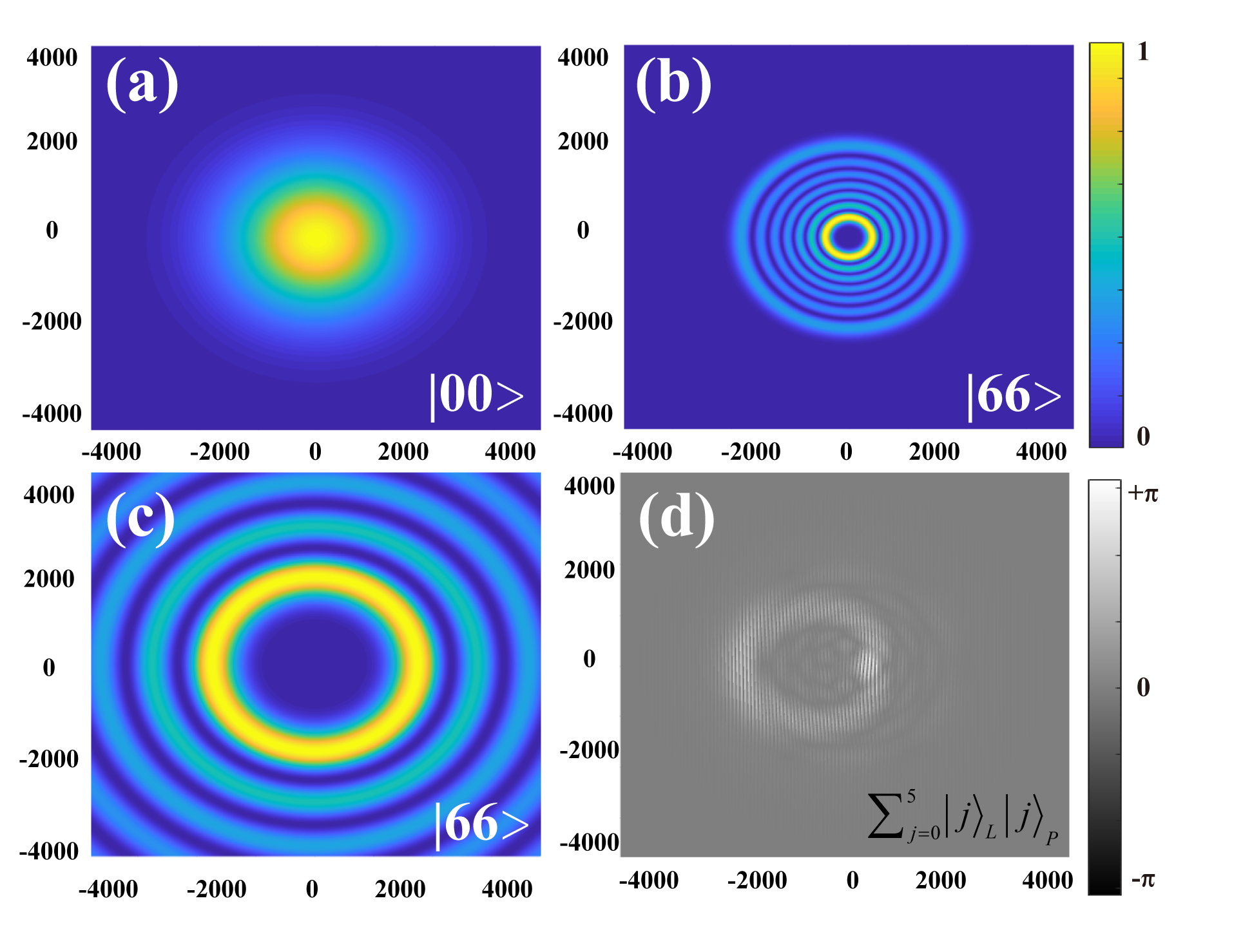}
  \caption{The intensity- and phase- distribution of the LG modes. (a): The intensity distribution of the fundamental Gaussian beam. (b) and (c): The intensity distribution of the LG modes ${\left| 6 \right\rangle _L}{\left| 6 \right\rangle _P}$ in standard and revised regimes, respectively. (d) The acquired phase for generation of six-dimensional maximally non-separable mode in SLM by the amplitude-phase and optical blazing encoding technology. In all the simulations, the wavelength and fundamental beam waist for a Gaussian beam are 780 nm and 2 mm.  }
\end{figure}

\subsection*{Simulation of mode orthogonality for angular--radius mode}
  By considering the Gaussian integral function of SMF, we simulate the orthogonality of standard and revised LG modes between the generation and detection. Fig. 4. (a) and 4(c) show the orthogonality for standard LG mode $LG_0^5$ and $LG_5^0$, where the detection modes are from $LG_0^0$ to $LG_0^{10}$, and $LG_0^0$ to $LG_{10}^0$, respectively. Fig. 4(b) and 4(d) are the situation of revised LG mode in our setup, where the beam waists for LG modes are manipulated by the relation of ${w_0}(L,P){\rm{ = }}{w_0}(0,0)/\sqrt {|L| + 2P + 1} $. For the standard LG mode, the orthogonality of radial mode gets worse due to the existing of GM-SMF, and it will obtain considerable improvement when choosing the revised LG modes. In our simulation, the size of mode $W$ equals to 750 um when it is imaged backward onto SLM, and the input Gaussian beam waist is equal to 1000 um.

Fig. 5 represents the distributions of intensity and phase of LG modes. Fig. 5(a) shows the fundamental Gaussian beam, where both the angular and radial number is zero. Fig. 5(b) is the intensity-distribution of LG modes (${\left| 6 \right\rangle _L}{\left| 6 \right\rangle _P}$) with the manipulation of beam waist in the above descriptions, and the Fig. 5(c) is the situation of the standard LG modes, where the beam size increases  dramatically as the radial and angular number. In this way [Fig. 5(c)], the amplitude-distributions of the state will exceed the size of screen in SLM for higher-order modes, which definitely gives rise to distortion for generation and detection. Our encoding regime that uses the revised LG mode can overcome the shortcoming [Fig. 5(b)].

Fig. 5(d) gives a hologram of six-dimensional angular-radial non-separable state ${\sum\nolimits_{j = 0}^5 {{{\left| j \right\rangle }_L}\left| j \right\rangle } _P}$, where the phase is encoded by amplitude-phase and optical blazing technology \cite{Bolduc2013,Liu20183DMES}. The regime can be expanded easily to the single photon level by changing the input source. Nevertheless, this type of encoding also exists the disadvantage, for example, the crosstalk will appear as the increase of radial mode, which is owing to the dislocation in many concentric rings between generation. In other words, the higher radial mode gets, the lower spatial resolution appears. An alternative method in experiment is to enlarge the size of the input beam.
\section*{Funding}
The Anhui Initiative in Quantum Information Technologies (AHY020200); National Natural Science Foundation of China (NSFC) (61435011, 61525504,  61605194); China Postdoctoral Science Foundation (2016M590570, 2017M622003); Fundamental Research Funds for the Central Universities.
\section*{Acknowledgments }
We thank Dr. Mehul Malik in Heriot-Watt University for helpful discussions in mode orthogonality.


\begin{thebibliography}{10}
\newcommand{\enquote}[1]{``#1''}

\bibitem{Einstein1935}
A.~Einstein, B.~Podolsky, and N.~Rosen, \enquote{Can quantum-mechanical
  description of physical reality be considered complete?}
  {\protect\JournalTitle{Phys. Rev.}} \textbf{47}, 777--780 (1935).

\bibitem{Clauser1978}
J.~F. Clauser and A.~Shimony, \enquote{Bell's theorem. experimental tests and
  implications,} {\protect\JournalTitle{Rep. Prog. Phys.}} \textbf{41}, 1881
  (1978).

\bibitem{Weihs1998}
G.~Weihs, T.~Jennewein, C.~Simon, H.~Weinfurter, and A.~Zeilinger,
  \enquote{Violation of bell's inequality under strict einstein locality
  conditions,} {\protect\JournalTitle{Phys. Rev. Lett.}} \textbf{81}, 5039
  (1998).

\bibitem{Collins2002}
D.~Collins, N.~Gisin, N.~Linden, S.~Massar, and S.~Popescu, \enquote{Bell
  inequalities for arbitrarily high-dimensional systems,}
  {\protect\JournalTitle{Phys. Rev. Lett.}} \textbf{88}, 040404 (2002).

\bibitem{Giovannetti2011}
V.~Giovannetti, S.~Lloyd, and L.~Maccone, \enquote{Advances in quantum
  metrology,} {\protect\JournalTitle{Nat. Photonics}} \textbf{5}, 222--229
  (2011).

\bibitem{Kok2007}
P.~Kok, W.~J. Munro, K.~Nemoto, T.~C. Ralph, J.~P. Dowling, and G.~J. Milburn,
  \enquote{Linear optical quantum computing with photonic qubits,}
  {\protect\JournalTitle{Rev. Mod. Phys.}} \textbf{79}, 135--174 (2007).

\bibitem{Kimble2008}
H.~J. Kimble, \enquote{The quantum internet,} {\protect\JournalTitle{Nature}}
  \textbf{453}, 1023--1030 (2008).

\bibitem{Gisin2002}
N.~Gisin, G.~Ribordy, W.~Tittel, and H.~Zbinden, \enquote{Quantum
  cryptography,} {\protect\JournalTitle{Rev. Mod. Phys.}} \textbf{74}, 145--195
  (2002).

\bibitem{Bouwmeester1997}
D.~Bouwmeester, J.-W. Pan, K.~Mattle, M.~Eibl, H.~Weinfurter, and A.~Zeilinger,
  \enquote{Experimental quantum teleportation,} {\protect\JournalTitle{Nature}}
  \textbf{390}, 575--579 (1997).

\bibitem{Spreeuw1998}
R.~J. Spreeuw, \enquote{A classical analogy of entanglement,}
  {\protect\JournalTitle{Found. Phys.}} \textbf{28}, 361--374 (1998).

\bibitem{Toeppel2014}
F.~T{\"o}ppel, A.~Aiello, C.~Marquardt, E.~Giacobino, and G.~Leuchs,
  \enquote{Classical entanglement in polarization metrology,}
  {\protect\JournalTitle{New J. Phys.}} \textbf{16}, 073019 (2014).

\bibitem{toninelli2019concepts}
E.~Toninelli, B.~Ndagano, A.~Vall{\'e}s, B.~Sephton, I.~Nape, A.~Ambrosio,
  F.~Capasso, M.~J. Padgett, and A.~Forbes, \enquote{Concepts in quantum state
  tomography and classical implementation with intense light: a tutorial,}
  {\protect\JournalTitle{Adv. Opt. Photonics}} \textbf{11}, 67--134 (2019).

\bibitem{doi:10.1080/00107514.2019.1580433}
T.~Konrad and A.~Forbes, \enquote{Quantum mechanics and classical light,}
  {\protect\JournalTitle{Contemp. Phys.}} \textbf{0}, 1--22 (2019).

\bibitem{aiello2015quantum}
A.~Aiello, F.~T{\"o}ppel, C.~Marquardt, E.~Giacobino, and G.~Leuchs,
  \enquote{Quantum- like nonseparable structures in optical beams,}
  {\protect\JournalTitle{New J. Phys.}} \textbf{17}, 043024 (2015).

\bibitem{Souza2007}
C.~Souza, J.~Huguenin, P.~Milman, and A.~Khoury, \enquote{Topological phase for
  spin-orbit transformations on a laser beam,} {\protect\JournalTitle{Phys.
  Rev. Lett.}} \textbf{99}, 160401 (2007).

\bibitem{lee2002experimental}
K.~Lee and J.~Thomas, \enquote{Experimental simulation of two-particle quantum
  entanglement using classical fields,} {\protect\JournalTitle{Phys. Rev.
  Lett.}} \textbf{88}, 097902 (2002).

\bibitem{Hasegawa2003}
Y.~Hasegawa, R.~Loidl, G.~Badurek, M.~Baron, and H.~Rauch, \enquote{Violation
  of a bell-like inequality in single-neutron interferometry,}
  {\protect\JournalTitle{Nature}} \textbf{425}, 45--48 (2003).

\bibitem{Zhou2016}
Z.~Zhou, Y.~Li, D.~Ding, W.~Zhang, S.~Shi, B.~Shi, and G.~Guo, \enquote{Orbital
  angular momentum photonic quantum interface,} {\protect\JournalTitle{Light:
  Science \& Applications}} \textbf{5}, e16019 (2016).

\bibitem{Fickler2012}
R.~Fickler, R.~Lapkiewicz, W.~N. Plick, M.~Krenn, C.~Schaeff, S.~Ramelow, and
  A.~Zeilinger, \enquote{Quantum entanglement of high angular momenta,}
  {\protect\JournalTitle{Science}} \textbf{338}, 640--643 (2012).

\bibitem{Karimi2010}
E.~Karimi, J.~Leach, S.~Slussarenko, B.~Piccirillo, L.~Marrucci, L.~Chen,
  W.~She, S.~Franke-Arnold, M.~J. Padgett, and E.~Santamato,
  \enquote{Spin-orbit hybrid entanglement of photons and quantum
  contextuality,} {\protect\JournalTitle{Phys. Rev. A}} \textbf{82}, 022115
  (2010).

\bibitem{Borges2010}
C.~Borges, M.~Hor-Meyll, J.~Huguenin, and A.~Khoury, \enquote{Bell-like
  inequality for the spin-orbit separability of a laser beam,}
  {\protect\JournalTitle{Phys. Rev. A}} \textbf{82}, 033833 (2010).

\bibitem{Goyal2013}
S.~K. Goyal, F.~S. Roux, A.~Forbes, and T.~Konrad, \enquote{Implementing
  quantum walks using orbital angular momentum of classical light,}
  {\protect\JournalTitle{Phys. Rev. Lett.}} \textbf{110}, 263602 (2013).

\bibitem{Cardano2015}
F.~Cardano, F.~Massa, H.~Qassim, E.~Karimi, S.~Slussarenko, D.~Paparo,
  C.~de~Lisio, F.~Sciarrino, E.~Santamato, and R.~W. Boyd, \enquote{Quantum
  walks and wavepacket dynamics on a lattice with twisted photons,}
  {\protect\JournalTitle{Sci. Adv.}} \textbf{1}, e1500087 (2015).

\bibitem{Perez-Garcia2018}
B.~Perez-Garcia, R.~I. Hernandez-Aranda, A.~Forbes, and T.~Konrad, \enquote{The
  first iteration of grover's algorithm using classical light with orbital
  angular momentum,} {\protect\JournalTitle{J. Mod. Opt.}} pp. 1--7 (2018).

\bibitem{DeOliveira2005}
A.~De~Oliveira, S.~Walborn, and C.~Monken, \enquote{Implementing the deutsch
  algorithm with polarization and transverse spatial modes,}
  {\protect\JournalTitle{J. Opt. B: Quantum Semiclassical Opt.}} \textbf{7},
  288--292 (2005).

\bibitem{Spreeuw2001}
R.~J. Spreeuw, \enquote{Classical wave-optics analogy of quantum-information
  processing,} {\protect\JournalTitle{Phys. Rev. A}} \textbf{63}, 062302
  (2001).

\bibitem{Kagalwala2013}
K.~H. Kagalwala, G.~Di~Giuseppe, A.~F. Abouraddy, and B.~E. Saleh,
  \enquote{Bell's measure in classical optical coherence,}
  {\protect\JournalTitle{Nat. Photonics}} \textbf{7}, 72--78 (2013).

\bibitem{Passos2018}
M.~Passos, W.~Balthazar, J.~A. de~Barros, C.~Souza, A.~Khoury, and J.~Huguenin,
  \enquote{Classical analog of quantum contextuality in spin-orbit laser
  modes,} {\protect\JournalTitle{Phys. Rev. A}} \textbf{98}, 062116 (2018).

\bibitem{dragoman2013quantum}
D.~Dragoman and M.~Dragoman, \emph{Quantum-classical analogies} (Springer
  Science \& Business Media, 2013).

\bibitem{Ndagano2017}
B.~Ndagano, B.~Perez-Garcia, F.~S. Roux, M.~McLaren, C.~Rosales-Guzman,
  Y.~Zhang, O.~Mouane, R.~I. Hernandez-Aranda, T.~Konrad, and A.~Forbes,
  \enquote{Characterizing quantum channels with non-separable states of
  classical light,} {\protect\JournalTitle{Nat. Phys.}} \textbf{13}, 397--402
  (2017).

\bibitem{Karimi2015}
E.~Karimi and R.~W. Boyd, \enquote{Classical entanglement?}
  {\protect\JournalTitle{Science}} \textbf{350}, 1172--1173 (2015).

\bibitem{berg2015classically}
S.~Berg-Johansen, F.~T{\"o}ppel, B.~Stiller, P.~Banzer, M.~Ornigotti,
  E.~Giacobino, G.~Leuchs, A.~Aiello, and C.~Marquardt, \enquote{Classically
  entangled optical beams for high-speed kinematic sensing,}
  {\protect\JournalTitle{Optica}} \textbf{2}, 864--868 (2015).

\bibitem{barreiro2008beating}
J.~T. Barreiro, T.-C. Wei, and P.~G. Kwiat, \enquote{Beating the channel
  capacity limit for linear photonic superdense coding,}
  {\protect\JournalTitle{Nat. Phys.}} \textbf{4}, 282--286 (2008).

\bibitem{marrucci2011spin}
L.~Marrucci, E.~Karimi, S.~Slussarenko, B.~Piccirillo, E.~Santamato, E.~Nagali,
  and F.~Sciarrino, \enquote{Spin-to-orbital conversion of the angular momentum
  of light and its classical and quantum applications,}
  {\protect\JournalTitle{J. Opt.}} \textbf{13}, 064001 (2011).

\bibitem{Salakhutdinov2012}
V.~Salakhutdinov, E.~Eliel, and W.~L{\"o}ffler, \enquote{Full-field quantum
  correlations of spatially entangled photons,} {\protect\JournalTitle{Phys.
  Rev. Lett.}} \textbf{108}, 173604 (2012).

\bibitem{loffler2012hybrid}
W.~L{\"o}ffler, V.~D. Salakhutdinov, and E.~R. Eliel, \enquote{Hybrid
  radial-angular quantum correlations of spatially entangled photons,} in
  \emph{Quantum Information and Measurement,}  (Optical Society of America,
  2012), pp. QW3A--6.

\bibitem{Allen1992}
L.~Allen, M.~W. Beijersbergen, R.~Spreeuw, and J.~Woerdman, \enquote{Orbital
  angular momentum of light and the transformation of laguerre-gaussian laser
  modes,} {\protect\JournalTitle{Phys. Rev. A}} \textbf{45}, 8185 (1992).

\bibitem{Dada2011}
A.~C. Dada, J.~Leach, G.~S. Buller, M.~J. Padgett, and E.~Andersson,
  \enquote{Experimental high-dimensional two-photon entanglement and violations
  of generalized bell inequalities,} {\protect\JournalTitle{Nat. Phys.}}
  \textbf{7}, 677--680 (2011).

\bibitem{Mair2001}
A.~Mair, A.~Vaziri, G.~Weihs, and A.~Zeilinger, \enquote{Entanglement of the
  orbital angular momentum states of photons,} {\protect\JournalTitle{Nature}}
  \textbf{412}, 313--316 (2001).

\bibitem{Yao2011}
A.~M. Yao and M.~J. Padgett, \enquote{Orbital angular momentum: origins,
  behavior and applications,} {\protect\JournalTitle{Adv. Opt. Photonics}}
  \textbf{3}, 161--204 (2011).

\bibitem{Karimi2014HOM}
E.~Karimi, D.~Giovannini, E.~Bolduc, N.~Bent, F.~M. Miatto, M.~J. Padgett, and
  R.~W. Boyd, \enquote{Exploring the quantum nature of the radial degree of
  freedom of a photon via hong-ou-mandel interference,}
  {\protect\JournalTitle{Phys. Rev. A}} \textbf{89}, 013829 (2014).

\bibitem{Krenn2017}
M.~Krenn, M.~Malik, M.~Erhard, and A.~Zeilinger, \enquote{Orbital angular
  momentum of photons and the entanglement of laguerre-gaussian modes,}
  {\protect\JournalTitle{Philosophical Transactions of the Royal Society A:
  Mathematical, Physical and Engineering Sciences}} \textbf{375}, 20150442
  (2017).

\bibitem{Trichili2016}
A.~Trichili, C.~Rosales-Guzmán, A.~Dudley, B.~Ndagano, A.~B. Salem, M.~Zghal,
  and A.~Forbes, \enquote{Optical communication beyond orbital angular
  momentum,} {\protect\JournalTitle{Sci. Rep.}} \textbf{6}, 27674 (2016).

\bibitem{Zhang2014}
Y.~Zhang, F.~S. Roux, M.~McLaren, and A.~Forbes, \enquote{Radial modal
  dependence of the azimuthal spectrum after parametric down-conversion,}
  {\protect\JournalTitle{Phys. Rev. A}} \textbf{89}, 043820 (2014).

\bibitem{Karimi2014}
E.~Karimi, R.~Boyd, P.~De~La~Hoz, H.~De~Guise,
  J.~{$\mathrm{\check{R}}$}eh{\'a}$\mathrm{\check{c}}$ek, Z.~Hradil, A.~Aiello,
  G.~Leuchs, and L.~L. S{\'a}nchez-Soto, \enquote{Radial quantum number of
  laguerre-gauss modes,} {\protect\JournalTitle{Phys. Rev. A}} \textbf{89},
  063813 (2014).

\bibitem{Zhang2018}
D.~Zhang, X.~Qiu, W.~Zhang, and L.~Chen, \enquote{Violation of a bell
  inequality in two-dimensional state spaces for radial quantum number,}
  {\protect\JournalTitle{Phys. Rev. A}} \textbf{98}, 042134 (2018).

\bibitem{Krenn2014}
M.~Krenn, M.~Huber, R.~Fickler, R.~Lapkiewicz, S.~Ramelow, and A.~Zeilinger,
  \enquote{Generation and confirmation of a (100$\mathrm{\times}$
  100)-dimensional entangled quantum system,}
  {\protect\JournalTitle{Proceedings of the National Academy of Sciences}} p.
  201402365 (2014).

\bibitem{Gu2018}
X.~Gu, M.~Krenn, M.~Erhard, and A.~Zeilinger, \enquote{Gouy phase radial mode
  sorter for light: Concepts and experiments,} {\protect\JournalTitle{Phys.
  Rev. Lett.}} \textbf{120}, 103601 (2018).

\bibitem{Zhou2017}
Y.~Zhou, M.~Mirhosseini, D.~Fu, J.~Zhao, S.~M.~H. Rafsanjani, A.~E. Willner,
  and R.~W. Boyd, \enquote{Sorting photons by radial quantum number,}
  {\protect\JournalTitle{Phys. Rev. Lett.}} \textbf{119}, 263602 (2017).

\bibitem{Bouchard2018}
F.~Bouchard, N.~H. Valencia, F.~Brandt, R.~Fickler, M.~Huber, and M.~Malik,
  \enquote{Measuring azimuthal and radial modes of photons,}
  {\protect\JournalTitle{Opt. Express}} \textbf{26}, 31925--31941 (2018).

\bibitem{Liu20183DMES}
S.~Liu, Z.~Zhou, S.~Liu, Y.~Li, Y.~Li, C.~Yang, Z.~Xu, Z.~Liu, G.~Guo, and
  B.~Shi, \enquote{Coherent manipulation of a three-dimensional maximally
  entangled state,} {\protect\JournalTitle{Phys. Rev. A}} \textbf{98}, 062316
  (2018).

\bibitem{Kues2017}
M.~Kues, C.~Reimer, P.~Roztocki, L.~R. Cort{\'e}s, S.~Sciara, B.~Wetzel,
  Y.~Zhang, A.~Cino, S.~T. Chu, and B.~E. Little, \enquote{On-chip generation
  of high-dimensional entangled quantum states and their coherent control,}
  {\protect\JournalTitle{Nature}} \textbf{546}, 622--626 (2017).

\bibitem{Li2018}
P.~Li, S.~Zhang, and X.~Zhang, \enquote{Classically high-dimensional
  correlation: simulation of high-dimensional entanglement,}
  {\protect\JournalTitle{Opt. Express}} \textbf{26}, 31413--31429 (2018).

\bibitem{Plick2015}
W.~N. Plick and M.~Krenn, \enquote{Physical meaning of the radial index of
  laguerre-gauss beams,} {\protect\JournalTitle{Phys. Rev. A}} \textbf{92},
  063841 (2015).

\bibitem{Bolduc2013}
E.~Bolduc, N.~Bent, E.~Santamato, E.~Karimi, and R.~W. Boyd, \enquote{Exact
  solution to simultaneous intensity and phase encryption with a single
  phase-only hologram,} {\protect\JournalTitle{Opt. Lett.}} \textbf{38},
  3546--3549 (2013).

\bibitem{Sephton2016}
B.~Sephton, A.~Dudley, and A.~Forbes, \enquote{Revealing the radial modes in
  vortex beams,} {\protect\JournalTitle{Appl. Opt.}} \textbf{55}, 7830--7835
  (2016).

\bibitem{vallone2017role}
G.~Vallone, \enquote{Role of beam waist in laguerre--gauss expansion of vortex
  beams,} {\protect\JournalTitle{Opt. Lett.}} \textbf{42}, 1097--1100 (2017).

\bibitem{Roux2014}
F.~S. Roux and Y.~Zhang, \enquote{Projective measurements in quantum and
  classical optical systems,} {\protect\JournalTitle{Phys. Rev. A}}
  \textbf{90}, 033835 (2014).

\bibitem{ndagano2018characterization}
B.~Ndagano and A.~Forbes, \enquote{Characterization and mitigation of
  information loss in a six-state quantum-key-distribution protocol with
  spatial modes of light through turbulence,} {\protect\JournalTitle{Phys. Rev.
  A}} \textbf{98}, 062330 (2018).

\bibitem{Liu2018Cat}
S.~Liu, Q.~Zhou, S.~Liu, Y.~Li, Y.~Li, Z.~Zhou, G.~Guo, and B.~Shi,
  \enquote{Generation of a macroscopic schr\" odinger cat using vortex light,}
  {\protect\JournalTitle{arXiv preprint arXiv:1807.05498}}  (2018).

\bibitem{Gibson2004}
G.~Gibson, J.~Courtial, M.~J. Padgett, M.~Vasnetsov, V.~Pas'ko, S.~M. Barnett,
  and S.~Franke-Arnold, \enquote{Free-space information transfer using light
  beams carrying orbital angular momentum,} {\protect\JournalTitle{Opt.
  Express}} \textbf{12}, 5448--5456 (2004).

\bibitem{Jack2010}
B.~Jack, A.~Yao, J.~Leach, J.~Romero, S.~Franke-Arnold, D.~Ireland, S.~Barnett,
  and M.~Padgett, \enquote{Entanglement of arbitrary superpositions of modes
  within two-dimensional orbital angular momentum state spaces,}
  {\protect\JournalTitle{Phys. Rev. A}} \textbf{81}, 043844 (2010).

\bibitem{kim2006phase}
T.~Kim, M.~Fiorentino, and F.~N. Wong, \enquote{Phase-stable source of
  polarization-entangled photons using a polarization sagnac interferometer,}
  {\protect\JournalTitle{Phys. Rev. A}} \textbf{73}, 012316 (2006).

\bibitem{hannam1999estimating}
M.~D. Hannam and W.~J. Thompson, \enquote{Estimating small signals by using
  maximum likelihood and poisson statistics,} {\protect\JournalTitle{Nucl.
  Instrum. Methods Phys. Res., Sect. A}} \textbf{431}, 239--251 (1999).

\bibitem{Reimer2018}
C.~Reimer, S.~Sciara, P.~Roztocki, M.~Islam, L.~R. Cort{\'e}s, Y.~Zhang,
  B.~Fischer, S.~Loranger, R.~Kashyap, and A.~Cino, \enquote{High-dimensional
  one-way quantum processing implemented on d-level cluster states,}
  {\protect\JournalTitle{Nat. Phys.}} \textbf{15}, 148--153 (2018).

\bibitem{acin2002quantum}
A.~Acin, T.~Durt, N.~Gisin, and J.~I. Latorre, \enquote{Quantum nonlocality in
  two three-level systems,} {\protect\JournalTitle{Phys. Rev. A}} \textbf{65},
  052325 (2002).

\bibitem{Cabello1996}
A.~Cabello, J.~Estebaranz, and G.~Garc{\'i}a-Alcaine,
  \enquote{Bell-kochen-specker theorem: A proof with 18 vectors,}
  {\protect\JournalTitle{Phys. Lett. A}} \textbf{212}, 183--187 (1996).

\bibitem{son2006generic}
W.~Son, J.~Lee, and M.~Kim, \enquote{Generic bell inequalities for multipartite
  arbitrary dimensional systems,} {\protect\JournalTitle{Phys. Rev. Lett.}}
  \textbf{96}, 060406 (2006).

\bibitem{datta2017measuring}
C.~Datta, P.~Agrawal, and S.~K. Choudhary, \enquote{Measuring
  higher-dimensional entanglement,} {\protect\JournalTitle{Phys. Rev. A}}
  \textbf{95}, 042323 (2017).

\bibitem{wang2017generation}
F.~Wang, M.~Erhard, A.~Babazadeh, M.~Malik, M.~Krenn, and A.~Zeilinger,
  \enquote{Generation of the complete four-dimensional bell basis,}
  {\protect\JournalTitle{Optica}} \textbf{4}, 1462--1467 (2017).

\bibitem{Babazadeh2017}
A.~Babazadeh, M.~Erhard, F.~Wang, M.~Malik, R.~Nouroozi, M.~Krenn, and
  A.~Zeilinger, \enquote{High-dimensional single-photon quantum gates: concepts
  and experiments,} {\protect\JournalTitle{Phys. Rev. Lett.}} \textbf{119},
  180510 (2017).

\bibitem{bavaresco2018measurements}
J.~Bavaresco, N.~H. Valencia, C.~Kl{\"o}ckl, M.~Pivoluska, P.~Erker, N.~Friis,
  M.~Malik, and M.~Huber, \enquote{Measurements in two bases are sufficient for
  certifying high-dimensional entanglement,} {\protect\JournalTitle{Nat.
  Phys.}} \textbf{14}, 1032--1037 (2018).

\bibitem{friis2018entanglement}
N.~Friis, G.~Vitagliano, M.~Malik, and M.~Huber, \enquote{Entanglement
  certification from theory to experiment,} {\protect\JournalTitle{Nat. Rev.
  Phys.}} \textbf{1}, 72--87 (2018).

\end{thebibliography}

\end{document}